\documentclass[conference]{IEEEtran}
\usepackage{cite}

\ifCLASSINFOpdf
  \usepackage[pdftex]{graphicx}
\else
\fi
\ifCLASSOPTIONcompsoc
  \usepackage[caption=false,font=normalsize,labelfont=sf,textfont=sf]{subfig}
\else
  \usepackage[caption=false,font=footnotesize]{subfig}
\fi
\usepackage{siunitx}

\begin{document}
%
\title{Entirely protecting operating systems against transient errors in space environment}

\author{
	\IEEEauthorblockN{
		Mahoukp\'{e}go Parfait Tokponnon \IEEEauthorrefmark{1} \IEEEauthorrefmark{2}\\ mahoukpego.tokponnon@uclouvain.be
		\and Marc Lobelle \IEEEauthorrefmark{1}\\ marc.lobelle@uclouvain.be 
		\and Eugene C. Ezin \IEEEauthorrefmark{2}\\eugene.ezin@ifri.uac.bj
	} \and 
    \IEEEauthorblockA{\IEEEauthorrefmark{1} Computing Science and Engineering Department\\
	Universit\'{e} catholique de Louvain \and 
	\IEEEauthorrefmark{2} Institut de Formation et de Recherche en Informatique \\Universit\'{e} d'Abomey-Calavi}%
}


%


\maketitle

\begin{abstract}
In this article, we propose a mainly-software hardening technique to totally protect unmodified running operating systems on COTS hardware against transient errors in heavily radiation - flooded environment like high altitude space. The technique is currently being implemented in a hypervisor and allows to control the upper layers of the software stack (operating system and applications). The rest of the system, the hypervisor, will be protected by other means, thus resulting in a completely protected system against transient errors. The induced overhead turns around 200\% but this is expected to decrease with future improvements.\\
\end{abstract}
\begin{IEEEkeywords}
	Transient errors, hypervisor, operating system, fault tolerance.
\end{IEEEkeywords}


%
\IEEEpeerreviewmaketitle

\section{Introduction}
A transient error is a change of state of a logical node in an electronic component (0 to 1 or 1 to 0) due to interaction between ionizing particles contained in cosmic rays and silicon atoms which generally compose integrated circuits. Although these errors do not damage the circuit, they may cause crashes, hangs and sometimes even erroneous results in Operating Systems (OS) and applications running on ordinary unprotected processors\cite{madeira2002experimental}. Therefore, for critical missions, ordinary equipment may not be used in an radiation-flooded environment without special care. Even though they are generally cheaper than hardware-hardened circuits  because the latter are manufactured in little series for niche markets.\\
In this article, we present a technique, still under research, that combines software redundancy with usual functionalities of ordinary hardware (blended technique) in order to totally protect operating systems running on these COTS\footnote{Commercial - Off - The - Shelf} materials against transient errors.
In the first part of this article, we present the technique in more detail, after specifying the objective of this work, then we will give some results we achieved so far.

\section{Blended hardening concept }
\subsection{Objective}
We propose to use some of the functionalities of ordinary hardware to detect and inhibit errors that occur in the circuit based on redundancy of execution. This is a Blended Hardening Technique (BHT) that allows to protect a full computing system at runtime with no need to access its source code.
\subsection{Blended hardening technique : Background}
\begin{figure}
	\centering
	\subfloat[Schematic view of PE execution]{%
		\includegraphics[width=.50\linewidth]{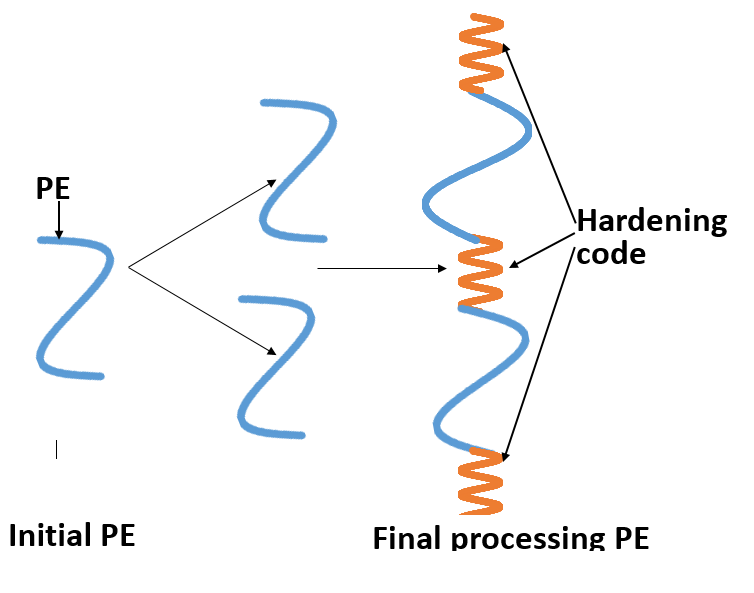}
		\label{fig:duplicate}
	}
	\subfloat[micro-hypervisor global architecture]{%
		\includegraphics[width=.50\linewidth]{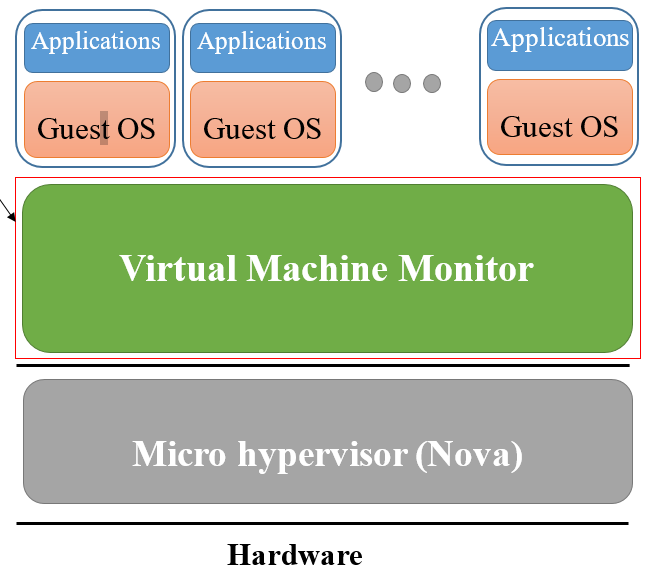}
		\label{fig:micro-hypervisor}
	}
	\caption{PE processing and Hypervisor view}	
\end{figure}

Considering a running program  which is a long stream of machine instructions, the BHT here consists of:
\begin{itemize}
	\item splitting, during execution, each program to be hardened into small sets of subsequent instructions called Processing Elements (PE), 
	\item running them twice,
	\item and comparing their execution traces to detect any occurring errors.
\end{itemize}
The Fig.\ref{fig:duplicate} gives a conceptual view of how each PE is processed. When errors are detected, the execution is simply rejected and the processing is resumed from the last correct execution point. The two execution have the same effect as the initial single execution. The PE is idempotent and its processing is atomic. For this to hold, the BHT sets two postulates:
\begin{itemize}
	\item Firstly, zero or at most one transient error per treatment: Thus, if there is an error, it would be either during the first execution, or during the second execution, or during the verification/commit phase, but never two or all phases will be erroneous. The statistical study in \cite{goka1991orbit} has shown that transient errors in an radiated environment follow the Poisson’s law. So, A maximum time interval, during which there can be no more than one transient error, can be deduced; no matter when the interval is taken.
    \item Secondly, a central memory, fully immune against transient errors, is necessary in order to preserve all data, coming from an error-free execution and saved in the main memory, from erroneous alteration. In this way one is sure to always start the processing of a PE from reliable data. The manner of obtaining such kind of memory from ordinary materials will not be discussed in this paper.
\end{itemize}

This model has been formally proven in \cite{lesage_software_2011}. In this article, such an interval had been calculated for a Leon processor (10\SI{}{\micro\second}) and a stand-alone program had been hardened by Lesage et al, using the BHT.\\
As the results were encouraging, the following step was to bring this technique to a more complex environment such as multitasking OS. This work is undergoing research where the kernel of Minix is being modified to harden its user applications that run on top of it \cite{assogba_etude_2011}. The hardening module being incorporated in the core, the OS itself remains thus unprotected. That's why we are searching to completely protect the OS, using a hypervisor (Nova \cite{steinberg_nova:_2010}) essentially because of its inherent ability to manipulate OS.

\section{Methodology}
\subsection{Hypervisor based hardening}
A hypervisor is software that runs directly on the hardware and can host one or more OS(s) in virtual machines. Thanks to the hypervisor, OSs run identically as if they were running on a bare machine (confer Fig. \ref{fig:micro-hypervisor}). Above the hypervisor and in user mode is the program called VMM (Virtual Machine Monitor) which is actually a set of system programs. Its role is to control, and emulate the virtual machine on which the guest system believes it is running.

\subsection{Approach}
To achieve this, we have subdivided the work into three steps:
\begin{itemize}
	\item Harden programs in the VMM
	\item Harden of OSs running on top of them
	\item Harden the hypervisor’s layer: but this last part is identical to the hardening of the standalone program Lesage had already achieved \cite{lesage_software_2011}. The hypervisor is also a standalone program that runs on the bare machine with the source code at disposal.
\end{itemize}

\section{Mid-term results} 
\begin{figure}[h]
	\centering
	\subfloat[Hardened run vs normal run]{%
		\includegraphics[width=.50\linewidth]{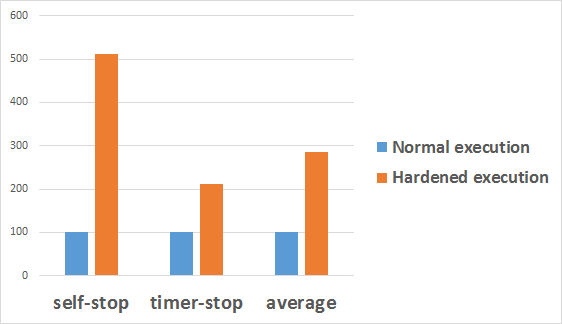}
		\label{fig:overhead-global}
	}
	\subfloat[Blended hardening vs Roman]{%
		\includegraphics[width=.50\linewidth]{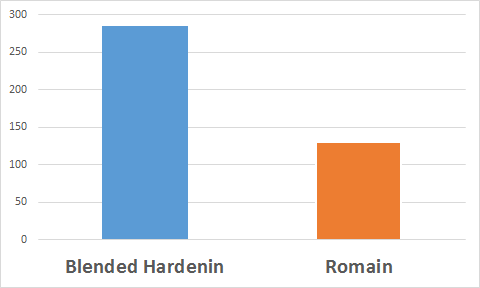}
		\label{fig:bht-romain}
	}
	\caption{BHT compared to normal execution and Romain}
\end{figure}

The first part of the work is currently finished, with the actual hardening of all the system processes in the VMM completely done. This means that an error that occurs during the execution of any VMM process is automatically detected and inhibited.
The overhead currently revolves around 2 times the normal execution in average. As shown in Fig \ref{fig:overhead-global}, this overhead is more severe when the hardened program release itself the CPU (self-stop) than when we had to interrupt it by a timer(timer-stop); that is because the PE execution time is shorter in the first case.
\section{Related works}
Other techniques have also been proposed in tis field. Although Romain \cite{dobel_operating_2014} ensures an overhead of 30\% when all redundant threads are spread on available CPU cores (Fig.\ref{fig:bht-romain}), it does not provide an entire protection against transient errors. We have not yet tested the BHT but this technique is designed to provide total protection for unmodified operating systems at runtime without needing to recompile them.
\section{Futurs works}
The rest of the work will be devoted specifically to the hardening of the virtualized OS. We will focus on:
\begin{itemize}
	\item privileged instructions management
    \item PE delimitation for guest OS and
	\item management of device drivers contained in guest OSs.
\end{itemize}
We will then test the system by both simulated transient errors and exposition to ionizing radiation in space environment to test it under real conditions.

\section{Conclusion}
We have shown in this paper a mainly-software method to protect the execution of operating systems on ordinary hardware against transient errors in highly radiation-flooded environment.\\We outlined the level reached and an overview of what remains to be done. Current level implementation gives a rather encouraging overhead and presages of acceptable overhead once the work will be finished.




\bibliographystyle{IEEEtran}
\bibliography{edcc}
%

\end{document}